\documentclass[lettersize,journal]{IEEEtran}
\usepackage{amsmath,amsfonts}
\usepackage{algorithmic}
\usepackage{algorithm}
\usepackage{array}
\usepackage[caption=false,font=normalsize,labelfont=sf,textfont=sf]{subfig}
\usepackage{textcomp}
\usepackage{stfloats}
\usepackage{url}
\usepackage{verbatim}
\usepackage{graphicx}
\usepackage{cite}

\begin{document}

\title{Functionalist Emotion Modeling in Biomimetic Reinforcement Learning}

\author{Louis Wang}



\maketitle

\begin{abstract}
We explore a functionalist approach to emotion by employing an ansatz—an initial set of assumptions—that a hypothetical concept generation model incorporates unproven but biologically plausible traits. From these traits, we mathematically construct a theoretical reinforcement learning framework grounded in functionalist principles and examine how the resulting utility function aligns with emotional valence in biological systems. Our focus is on structuring the functionalist perspective through a conceptual network, particularly emphasizing the construction of the utility function, not to provide an exhaustive explanation of emotions. The primary emphasis is not on planning or action execution, but such factors are addressed when pertinent. Finally, we apply the framework to psychological phenomena such as humor, psychopathy, and advertising, demonstrating its breadth of explanatory power.
\end{abstract}


\section{Introduction}
\IEEEPARstart{A}{n} ansatz is a German term meaning "approach" or "starting point," commonly used in physics, mathematics, and related fields. It refers to an educated guess or assumption about the form of a solution to a problem, particularly when solving complex equations or modeling physical systems. By making specific assumptions about the structure or behavior of a system, researchers can simplify intricate problems into more manageable forms. For example, in quantum mechanics, an ansatz might involve proposing a particular wavefunction form for a particle, while in statistical physics, it could involve assuming equilibrium conditions for a system. While an ansatz introduces simplifications, it is typically based on known physical principles or empirical observations to ensure its validity. This approach allows scientists and engineers to derive solutions that capture the essential features of a problem without getting bogged down by overwhelming complexity.

Understanding biological brains, especially emotional and learning processes, exemplifies such intricate challenges. At present, the fundamental set of mechanisms governing how biological brains learn remains unclear. Drawing parallels between current machine learning paradigms and human cognition serves as a valuable tool but also has its limitations. While contemporary machine learning systems have achieved remarkable results, there are notable differences between these systems and human cognition in critical areas such as sparsity, data efficiency, resilience to adversarial attacks, and handling out-of-distribution inputs \cite{Lake_2016, Geirhos_2018}. Specifically, research in artificial neural network (ANN) reinforcement learning tends to prioritize outputs over understanding internal states, largely because of the enduring challenge posed by opaque, high-dimensional hidden layers \cite{Angelov_2021}. In contrast, biological systems have identifiable internal emotional states, demonstrated even in relatively simple processes such as Pavlovian conditioning, where overt observable behavioral outputs are not necessary \cite{Neftci_2019}. Thus, while simple, model-free reinforcement learning is well-documented from both neurobiological and computational perspectives, the mechanisms behind model-based reinforcement learning in biological systems, including internal emotional states, remain poorly understood.

\section{The Ansatz}

To make progress bridging the gap between biological neural networks and ANNs, we attempt an ansatz, whereupon we make certain assumptions about a hypothetical learning mechanism to more closely resemble the biological system. We aim for a set of assumptions that are minimal, functional, and plausible. Specifically, we assume the following:

\begin{enumerate}
\item[(i)] Learning is unsupervised and builds from the bottom up.

\item[(ii)] Sparsity. In particular, active concepts represented by subset of active perceptrons with binary activation (1 or 0, all or nothing), and perceptrons representing inactive concepts remain inactive.

\item[(iii)] Hierarchy of concepts across layers.
\end{enumerate}

\subsection{Rationale of assumptions}

The current framework for learning in artificial neural networks relies on a top-down gradient descent approach, where an error function is minimized through iterative processing of training data \cite{Rojas_1996}. In contrast, biological neural networks exhibit a more complex and less transparent learning mechanism, with evidence suggesting that they employ both bottom-up and top-down processes. For our purposes, assumption (i) restricts the learning process to an unsupervised, bottom-up approach, which is acknowledged as a fundamental component of the broader learning framework \cite{Ghose_2004, Kourtzi_2006}. 

Assumption (ii) is counterintuitive from an information density perspective. Given an n-bit bandwidth, more data can be represented if each permutation is allowed to represent a separate concept. However, this assumes that maintaining a 1 or 0 state requires the same energy. In biological systems, this is simply not true, hence the preference for sparsity \cite{Lennie_2003}. From the analysis offered in Levy and Baxter \cite{Levy_1996}, the cost of 10 spikes in 200 ms is more than 100 times that of maintaining the resting potential, and under those conditions the optimal fraction of active neurons is about 3\%. Given this disparity, we can see how an encoding scheme satisfying assumption (ii) may increase evolutionary fitness, especially given the high energy cost of brain mass even with such optimizations.

The final assumption (iii) is a trait that is observed in biological systems \cite{Ghose_2004, Kourtzi_2006}, and to varying degrees observed in artificial neural networks trained using backpropagation \cite{Krizhevsky_2012} and more biologically plausible approaches such as spike timing dependent plasticity \cite{Masquelier_2007}.

From these assumptions, we construct a reinforcement learning framework using the functionalist approach \cite{Adolphs_2018}. This lens considers emotions as systems evolved to address evolutionary fitness, which will be reflected in our design decisions.

\section{Initial State}

When the network is first initialized, we assume that no features have been learned, and all perceptrons in the network are plastic and unspecialized. Under these conditions, the constructionist view espoused by Barrett \cite{Barrett_2017} is not yet useful- after all, how can there be constructed emotions when no constructs have been learned? Instead, the main information available is raw sensory input, and the basic behavioral directives are avoidance and approach (positive and negative valence, with the exception of anger discussed later). Evolution or genetics can define the initial state of the system, including the valence of specific sensory inputs. Chemical sensors on the tongue detecting a sweet substance can yield a positive valence output, while something sour can yield a negative valence. And depending on the specific sensory valence activated, generally appropriate automatic responses can be defined. For example, high negative valence in taste implies an inedible object, thus gagging and retching reflexes are attached to the given sensory activations.

These basic reactions are well-understood biologically and well-modeled computationally, so we will abstain from more detail. For the sake of convenience, we will name this initially-defined utility function based on raw sensory input the primary utility function (P-UF). This may correspond to the classical or basic view of emotion, where they are universal across cultures \cite{Ekman_1992, MacLean_1970}.

Finally, while perceptrons in hidden layers are initially plastic and unspecialized, they can be assigned an initial utility value unbound to any P-UF component. If a positive valence is associated with the activation of these perceptrons, behaviors that activate them are more likely to be approached, potentially mirroring the concept of curiosity and its role in reinforcing exploratory behavior.

\section{Emotional Constructs}

As stated before, the uninitialized perceptrons are relatively neutral. How then are more advanced emotional constructs described by Barrett \cite{Barrett_2017} created? Due to the unsupervised nature of the learning algorithm, we assume that the concepts themselves can be generated through relatively passive observations of the environment through all perceivable input modalities without the need for emotional drive, though emotions can certainly help or hinder the process (e.g. curiosity or depression, respectively). However, the neutrality ends given P-UF activity- all simultaneously active perceptrons representing all active concepts (assumption ii) are mapped with the corresponding valence of the active P-UF components, with the magnitude of the mapping roughly proportional to the magnitude of the P-UF valence. In other words, active concepts (via assumption ii) gain emotional value from sensory responses. Persistence of biological emotion as well as memory and priming effects may result in valence mapping onto perceptrons active shortly before and after initial P-UF activation. The associations between active concepts may be strengthened in proportion to the magnitude of valence. With mapped P-UF valence, these composite concepts appear to become emotional constructs, which guide approach or avoidance behaviors. For the sake of convenience, we will name the total collection of learned emotional constructs the secondary utility function (S-UF).

\subsection{Functional differences between the primary and secondary}

From a functional or evolutionary standpoint, why do these additional emotional constructs exist beyond the P-UF? Because the P-UF is limited to assessing the evolutionary significance of immediately noticeable events, additional emotional constructs are needed to predict significant emotional experiences. Through these predictors, the organism is able to act accordingly by seeking rewards or avoiding harm before such outcomes materialize.

As P-UF serves as an anchor for assessing the utility of immediate, materialized events, its components cannot be extinguished. Instead, they can only be exhausted under specific evolutionary purposes. Positive valence components, for instance, can be depleted to prevent reward hacking and other behaviors that may have biological costs, such as overeating. In contrast, negative valence components do not face the same risk of reward hacking due to their avoidance-driven nature, allowing them to persist more readily. However, negative S-UF components can still eventually be depleted, unlike the primary counterparts. To use an analogy, if P-UF is a "valence faucet," S-UF is the "cup" it fills, with a finite, conserved local resource.

Once a primary component is exhausted or satiated, the associated elements of the S-UF no longer necessitate continued approach or avoidance. This function suggests a persistent link between constructed S-UF components and the P-UF components integral to their formation, establishing a framework through which relevant S-UF components can be identified for targeted suppression or excitation as needed, without affecting the mapped valence. Without this link, hunger may mistakenly trigger S-UF components linked to oxygen deprivation, driving behavior that is irrelevant to immediate needs. In addition, this link also establishes a mechanism by which an activated S-UF component can trigger the arousal symptoms of the P-UF components integral in its creation.

\subsection{S-UF Valence Transfer}

While the P-UF has a global effect, S-UF can spread in a local manner, limited by the valence available locally. S-UF valence transfer occurs when activation of an S-UF component is involved in activating perceptrons outside the component, modeled by a diffusion process we formulate in later sections. As suggested previously, S-UF valence is conserved, barring inefficiencies or P-UF involvement. 

\subsection{Extinction}

The function of the S-UF is to identify and predict its associated P-UF components through learned experiences. However, due to the complexity of reality, these learned associations may not always be accurate. The question arises: how are incorrect associations eliminated and correct associations maintained? In contrast to P-UF, we define the S-UF such that depleted valence is not restored over time by default. If a positive S-UF component is repeatedly activated without concurrent activation of its associated P-UF component, the S-UF component will deplete entirely, leading to extinction. Conversely, if the activation of the P-UF component follows the activation of the S-UF component, the mapped valence of the S-UF component is restored through the same process that created it.

In the context of negative valence S-UFs, the dynamics differ significantly compared to their positive counterparts. Unlike positive valence, the activation of a predicted negative valence P-UF component is not the desired behavior, avoidance is. Consequently, relying on P-UF activation to sustain the S-UF mapping is an inherently risky strategy. If negative valence S-UF components deplete or extinguish too rapidly for a robust evasion motor output strategy to develop, or if an established evasion response deteriorates prematurely, the associated harmful outcome is likely to recur. Therefore, it is essential to slow the depletion and extinction rates of negative S-UF components sufficiently to allow the formation of a robust evasion response, and the response itself must be resistant to degradation over time. All objectives can be achieved by increasing the magnitude and duration of negative P-UF components, which allows more negative valence to be mapped to the corresponding S-UF component during the initial experienced association. With increased local valence, the negative S-UF component takes longer to deplete, which in turn provides more time for the avoidance strategy to be reinforced effectively. This is observed experimentally by Solomon et al. \cite{Solomon_1953}.

Overall, the distinction between primary and secondary utility functions highlights the difference between emotional states associated with anticipating an outcome versus those experienced as a reaction to the outcome itself. In the context of positive valence, this separation differentiates between craving—anticipating a pleasurable experience—and savoring—the enjoyment derived from the actual experience. Similarly, for negative valence, it distinguishes fear, which involves anticipation of harm or danger, from pain, which is the distress experienced as a direct result of an adverse outcome.

\subsection{Opposite Valence Interactions}

Another important consideration is the interaction between positive and negative originating from both primary and secondary sources. A straightforward approach to address this issue is to utilize net valence. This method is generally acceptable during the initial development of an S-UF component, as costs are often associated with obtaining rewards in such contexts. For instance, enjoyment of spicy foods is not unusual, but enjoyment of the spice in isolation is. Additionally, employing net value implies that when positive and negative valences occur simultaneously within the same construct, they will offset each other.

In the context of natural environments, sources of reward hold significant value. The net valence approach reveals that when a mature, previously learned positive S-UF component co-activates with an active negative P-UF or S-UF, it leads to the erosion of the positive source. This behavioral response is advantageous for prey, which primarily rely on avoidance. However, in scenarios where defense is more appropriate, this approach may not suffice. In cases where the positive S-UF components suffer a significant decrease, a new marker is mapped onto currently active negative valence sources. This process temporarily protects the positive source from erosion, providing time for outward-facing actions to address the threat effectively. In everyday terms, this response can be described as anger. For convenience, let us label the component responsible for planning and execution of outward-facing actions as the Decision-Execution (DE) module, which takes P-UF and S-UF responses as inputs to decide on courses of action.

Due to the relatively untargeted nature of unmodulated anger, it is also possible that the sources of negative valence are not specifically targeted initially. Rather, all active perceptrons representing active concepts outside of the threatened positive S-UF component are. Given that design, initial anger has a more global effect akin to that of P-UF activation. Concepts can then be marked by the initial anger and diffusion, similar to S-UF construction. The DE module can modulate further through mechanisms such as appraisal and attention, resulting in a more targeted response.

The activation of a S-UF component without corresponding P-UF activation also represents a potential threat, albeit not stemming from external negative valence sources. Instead, this threat arises due to the inability of the DE module to obtain predicted rewards. Similar to an anger reaction, the S-UF component can be temporarily shielded from extinction, and its mapped positive valence amplified, to provide time for and further motivate the DE module to achieve its objectives. However, should these temporary measures fail to induce associated P-UF activation, the S-UF component will eventually deplete and become extinguished, as it does under previous conditions. If negative valence is used in conjunction with the anger marker, then this situation aligns with the observed outcomes of frustration by Amsel \cite{Amsel_1992}.

\subsection{Targeting P-UF valence outputs}

Under assumption (iii), a hierarchical structure of features across layers is typically observed. If the distribution of concepts across these layers can be somewhat predicted, the global effect of the P-UF may not remain entirely uniform. Instead, the valence output targets associated with the P-UF could be weighted through evolutionary tuning according to the likelihood of their relevance to the corresponding component within a given layer. For instance, in a feed-forward visual processing network, the emphasis on mapping valence might shift away from lower-level features, such as edge orientation, and toward higher-order concepts, like object recognition, where their contribution to the P-UF is more significant.

Valence output targets should generally align with the modality of their associated P-UF components within a collection of unimodal feed-forward networks. However, biological neural networks often incorporate regions where multiple modalities are integrated. Language comprehension, for instance, is a prominent multimodal domain that generates robust representations. Consequently, it would not be unexpected if language-processing areas, such as Wernicke's area, were disproportionately emphasized by most, if not all, P-UF components in their operations.

To examine a concrete case under the functionalist framework, consider the olfactory P-UF components and their role in sensory perception. Olfaction is closely tied to gustation, serving as a predictor of edible goods—a function that aligns with its evolutionary purpose. Additionally, the distribution of smells follows an approximate inverse-square law, meaning odor intensity increases significantly as one approaches the source. Given that humans are inherently social creatures who also emit scents, this creates opportunities for fostering interpersonal focus. A positive P-UF component could be designed to detect human scent while outputting valence at object-level concepts, encouraging attention toward other individuals. Furthermore, since early mammals typically have heightened reliance on olfactory cues compared to visual ones \cite{Heesy_2010}, reducing the input signal to the olfactory modality through the evolutionary process may shift olfactory-driven valence toward associated visual concepts. These speculations align with empirical observations that, relative to males, human females exhibit greater sensitivity to smells \cite{Doty_1984} and demonstrate stronger interest in people over objects \cite{Connellan_2000} even before social conditioning can take place.

In addition, persistent P-UF to S-UF links are supported by physiological responses such as physical disgust or salivation underpinning avoidance or approach behaviors tied to these social concepts. In this model, visceral disgust elicited by negative social behavior is hypothesized to originate from an olfactory-based P-UF. While existing research \cite{Rozin_2008, Wicker_2003} acknowledges olfaction’s evolutionary role in disgust and its extension to social contexts via neural pathways (e.g., insula activation), it typically frames these effects as associative or descriptive rather than utility-driven. Similarly, studies like Stevenson and Repacholi \cite{Stevenson_2005} highlight olfaction’s impact on social preferences, but they do not emphasize a hierarchical, persistent utility transfer.

\subsection{Planning and modulation}

In real-world scenarios, situations often involve valences that fluctuate over time, with short-term outcomes conflicting with long-term consequences. These temporal variations necessitate planning beyond a simple greedy approach when the long-term valence is not only greater but also opposite to the short-term valence. During the execution of such plans, the DE module must operate with at least temporary insulation from the immediate valence, ensuring that long-term objectives can be pursued effectively without being derailed by transient feedback. This aligns with the role of the prefrontal cortex in strategic planning and its ability to regulate emotional responses \cite{Gross_1998, Miller_2001}. Future refinements could explore attention as a valence redirection tool within the DE module.

\subsection{Evolutionarily Primed S-UF}

Evolutionary evidence suggests that beyond simple reflexes, preprogrammed behaviors can emerge through evolutionary processes, as seen in examples such as the stereotyped caching behavior in squirrels \cite{Preston_2009}. Similarly, raw sensory input may be processed to classify evolutionarily significant concepts, as in the case of aposematism \cite{Rowe_1999} and mate selection \cite{Verzijden_2007}. However, most if not all such responses must be acquired- in other words, the concept class is easily learned and can readily acquire significant valence via higher perceptron valence capacity or strong P-UF weights, but is valence-neutral upon creation. These S-UF concepts are easily learned due to hardwired aids like the fusiform face area for facial cues, and heavily weighted by P-UF valence outputs for rapid, potent mapping. In social animals like humans, this class excels in social dimensions, enabling swift emotional tagging of cues critical for survival and bonds. It retains S-UF’s learned flexibility while leveraging innate support for valence acquisition critical for evolutionary fitness. However, this flexibility is double-edged and can turn maladaptive, which we will explore in a later section.

\section{Formulations}

The model's mathematical framework is constructed around a perceptron network with two primary parts: the Primary Utility Function (P-UF) and the Secondary Utility Function (S-UF). These parts operate under assumptions (i-iii), with P-UF components providing immediate valence from sensory inputs and S-UF perceptrons learning emotional associations. Below, we define these parts and outline a set of rules modeling valence dynamics.

\subsection{Definitions}

\begin{enumerate}
    \item[(i)] \textbf{P-UF Components}: A set \(\{P_i \mid i \in \mathbb{N} \} \), where a single component is represented as \( P_i \) (\( i \) indexes the components). Each is activated by specific sensory inputs (e.g., a sweet taste or a loud noise). When active, \( P_i \) outputs a valence \( v_i(t) \), defined as:
    \begin{equation}
    v_i(t) = v_i \cdot f_p(P_i, t)
    \end{equation}
    where \( v_i \) is the intrinsic valence of \( P_i \), and \( f_p(P_i, t) = 1 \) if active at time \( t \), 0 otherwise.

    \item[(ii)] \textbf{S-UF Perceptrons}: A set \(\{s_i \mid i \in \mathbb{N}\}\), where a single perceptron is represented as \( s_i \) (where \( i \) indexes the perceptrons). Each has a binary activation state \( f_s(i, t) = 1 \) if active at time \( t \), 0 otherwise, and a utility value \( UV_i(t) \) representing learned valence. A single S-UF component is composed of a subset of all perceptrons, \(S_i \subseteq \{s_j \mid j \in \mathbb{N}\}\)

    \item[(iii)] \textbf{Eligibility Trace}: Each perceptron maintains an eligibility trace \( e_i(t) \), reflecting recent activity and providing temporal flexibility:
    \begin{equation}
    e_i(t) = \alpha \cdot e_i(t-1) + (1-\alpha) \cdot f_s(i, t)
    \end{equation}
    where \( \alpha \) (\( 0 < \alpha < 1 \)) is a decay factor.

    \item[(iv)] \textbf{Valence}: A scalar quantity that can be positive or negative, reflecting reward or punishment.

    \item[(v)] \textbf{Time}: The system evolves in discrete time steps \( t \), allowing for temporal updates.
\end{enumerate}

\subsection{P-UF to S-UF Valence Mapping}

P-UF components distribute valence to S-UF perceptrons based on their eligibility trace, enabling temporal flexibility in emotional reinforcement. The valence input to S-UF perceptron \( s_i \) from all P-UF components at time \( t \) is:

\begin{equation}
PUV_i(t) = \sum_j w_j(i) \cdot v_j(t) \cdot e_i(t)
\label{eq:puv}
\end{equation}
where \( w_j(i) \) is the evolutionarily tuned weight representing the association strength between P-UF component \( P_j \) and S-UF perceptron \( s_i \). This formulation ensures that perceptrons recently active (with \( e_i(t) > 0 \)) can receive valence, reflecting persistent emotional effects.

\subsection{Collective Valence of S-UF Components}

An S-UF component \( S_k \) (where \( k \) indexes distinct concepts) is defined as a subset of perceptrons \( S_k \subseteq \{s_j \mid j \in \mathbb{N}\} \), representing a coherent concept. While the network operates at the perceptron level, emotional valence is often experienced and interpreted at the component level. To reconcile these perspectives, we define the collective valence of an S-UF component as follows.

The \textbf{collective valence} of \( S_k \) at time \( t \), denoted \( UV_{S_k}(t) \), represents the emotional value of the concept based on its currently active perceptrons:
\begin{equation}
UV_{S_k}(t) = \sum_{s_i \in S_k} UV_i(t) \cdot f_s(i, t)
\label{eq:partial_activation}
\end{equation}
Here, only perceptrons active at time \( t \) (\( f_s(i, t) = 1 \)) contribute to the component’s valence. This formulation captures \textit{partial activation}: the valence experienced depends on how many perceptrons within \( S_k \) are active, reflecting varying emotional intensity.

For comparison, the \textbf{total component valence}, or maximum potential valence of \( S_k \) if all perceptrons were active, is:
\begin{equation}
UV_{S_k}^{\text{max}} = \sum_{s_i \in S_k} UV_i(t)
\label{eq:total_valence}
\end{equation}
This distinguishes the current emotional experience from the concept’s full emotional potential.

\subsection{S-UF Valence Dynamics}

S-UF perceptrons evolve through two internal processes: decay and diffusion with transfer efficiency.

\textbf{\textit{Diffusion with Transfer Efficiency:}} Valence transfers between connected S-UF perceptrons (denoted \( j \sim i \)) with a loss factor \( \eta \) (\( 0 \leq \eta \leq 1 \)), modeling potential loss during emotion diffusion. The change in utility due to diffusion in perceptron \(i\) is \(d_i(g, t)\), where \(g(i, t)\) is a function tracking some local resource of perceptron \(i\) at time \(t\), and \(d_i\) defined as the following:
\begin{equation}
d_i(g, t) = \beta \cdot \sum_{j \sim i} ((g(j, t) - g(i, t)) \cdot f_s(j, t)
\end{equation}

where \( \beta \) (\( 0 < \beta < 1 \)) is the diffusion rate. The transfer loss \(l_i(t)\) is defined as follows:
\begin{equation}
l_i(g, t)=\text{sign}(g(i, t)) \cdot \eta \cdot \sum_{j \sim i} |g(j, t) - g(i, t)| \cdot f_s(j, t)
\end{equation}

where \(\eta\) (\( 0 \le \eta \leq 1 \)) is the loss rate. When \( \eta > 0 \), a fraction of the valence is lost, reducing total valence in the system. The total diffusion term is:
\begin{equation}
\text{diffusion}_i(t) = f_s(i, t) \cdot (d_i(g_\text{uv}, t) - l_i(g_\text{uv}, t))
\label{eq:diffusion_valence}
\end{equation}

where \(g_\text{uv}(i, t) = UV_i(t)\)

\subsection{Utility Update Rule}

The utility value \( UV_i(t) \) of each S-UF perceptron is updated by combining previous state, P-UF valence input, and diffusion:
\begin{equation}
UV_i(t+1) = UV_i(t) + PUV_i(t) + \text{diffusion}_i(t)
\label{eq:uv_update}
\end{equation}

To prevent unbounded growth, we clip the utility values:
\begin{equation}
UV_i(t+1) = \text{clip}(UV_i(t+1), -L_i, L_i)
\label{eq:p_valence_limit}
\end{equation}
where \( L_i \) is a perceptron-specific limit on valence magnitude.

\subsection{Properties}

\begin{itemize}
    \item \textbf{Reinforcement}: When P-UF components activate, \( PUV_i(t) \) replenishes or adjusts \( UV_i(t) \) based on recent activity via \( e_i(t) \), counteracting depletion and strengthening associations.
    \item \textbf{Depletion}: Without P-UF support, active S-UF perceptrons lose valence through inefficient diffusion (\( \eta > 0 \)).
    \item \textbf{Valence Transfer}: The diffusion term spreads valence locally among active, connected perceptrons, with \( \eta \) controlling efficiency. If \( \eta = 0 \), transfer is conservative; if \( \eta > 0 \), some valence is lost.
    \item \textbf{Temporal Flexibility}: The eligibility trace \( e_i(t) \) ensures that P-UF valence influences S-UF perceptrons active shortly before or during activation, aligning with persistent emotional effects.
\end{itemize}

The global effect of P-UF is captured by distributing \( v(t) \) to active S-UF perceptrons via weights, with eligibility traces allowing for temporal flexibility. The diffusion term enables local valence spread among S-UF perceptrons, supporting emotional association and opposite valence annihilation. Parameters such as \( \alpha \), \( \beta \), \( \eta \), and \( L_i \) can be tuned to reflect decay factor, diffusion rate, transfer inefficiencies, and perceptron-level valence capacity, aligning with biological emotion and memory effects.

\subsection{Anger Dynamics}

Anger emerges as a protective mechanism within the network, triggered by significant losses in positive valence, and serves to defend positive S-UF components through reduced diffusion and limited valence recovery. We outline a mathematical formulation of anger dynamics below.

\subsubsection{Local and Global Valence Tracking}
For each perceptron \( s_i \), an \textit{exponential moving average} (EMA) tracks its baseline valence:
\begin{equation}
\text{EMA}_i(t) = \gamma \cdot \text{EMA}_i(t-1) + (1 - \gamma) \cdot UV_i(t)
\end{equation}
where \( \gamma \) (\( 0 < \gamma < 1 \)) is the smoothing factor, and \( UV_i(t) \) is the current utility value. A global baseline sums the EMA of all perceptrons with positive valence:
\begin{equation}
\text{Baseline}(t) = \sum_{i : \text{EMA}_i(t) > 0} \text{EMA}_i(t)
\end{equation}
The current total positive valence is:
\begin{equation}
\text{CurrentPositive}(t) = \sum_{i : \text{EMA}_i(t) > 0} UV_i(t)
\end{equation}

\subsubsection{Anger Activation}
Anger \( A(t) = A_p(t) + A_s(t)\), where \(A_p(t)\) is the global/primary anger term and \(A_s(t)\) is the local/secondary anger term. \(A_p(t)\) activates when the loss of positive valence exceeds a threshold \( \theta \):
\begin{equation}
\text{Loss}(t) = \text{Baseline}(t) - \text{CurrentPositive}(t)
\end{equation}
\begin{equation}
A_p(t) = \text{max}\left(\text{Loss}(t) - \theta, 0\right) + \alpha_a \cdot A_p(t-1)
\label{eq:global_anger}
\end{equation}
where \(0 < \alpha_a < 1\) is the decay factor for global anger. \(A_s(t)\) is the sum of local anger scalars \(a_i(t)\) among all active perceptrons:
\begin{equation}
A_s(t) = \sum_{i \in \mathbb{N}} a_i(t) \cdot f_s(i, t)
\label{eq:local_anger}
\end{equation}

Primary anger decays naturally over time. As the \(\text{Baseline}(t)\) normalizes, \(\text{Loss}(t)\) decreases, driving \(A_p(t)\) lower.

\subsubsection{Defensive Mechanisms}
Anger triggers two defensive effects on positive S-UF components:
\begin{itemize}
    \item \textbf{Reduced Valence Diffusion}: The diffusion rate \( \beta \) as related to normal UV flow adjusts dynamically, limiting diffusion as anger increases:
    \begin{equation}
    \beta_{\text{effective}}(t) = \beta \cdot \left(1 - \frac{\text{min}(A(t), A_\text{thres})}{A_{\text{thres}}}\right)
    \label{eq:beta_effective}
    \end{equation}
    where \(A_\text{thres}>0\) is the threshold of total anger beyond which P-UF inputs and normal UV diffusion inputs are frozen.
    \item \textbf{Limited Recovery of Lost Valence}: For perceptrons where \( \text{EMA}_i(t) > UV_i(t) \) and \( \text{EMA}_i(t) > 0 \), anger facilitates a limited recovery process.
\end{itemize}

\subsubsection{Limited Recovery Mechanism}
To prevent unbounded valence growth, we track the total recovered utility value per perceptron using an \textit{exponential moving sum} (EMS):
\[
\text{EMS\_recovery}_i(t) = \alpha \cdot \text{EMS\_recovery}_i(t-1) \]
\begin{equation}
+ \text{recovery}_i(t)
\end{equation}
where \( \alpha \) (\( 0 < \alpha < 1 \)) is the decay, and \( \text{recovery}_i(t) \) is the recovery applied at time \( t \), with \( \text{EMS\_recovery}_i(0) = 0 \). The recovery cap is:
\begin{equation}
\text{cap}_i(t) = k \cdot \text{EMA}_i(t)
\end{equation}
where \( k \ge 0 \) (e.g., 1.5) is a factor ensuring the cap exceeds the baseline. The maximum allowable recovery is:
\[
\text{max\_recovery}_i(t) = \text{clip}(\sigma + \text{max\_recovery}_i(t-1)\]
\begin{equation}
- \text{EMS\_recovery}_i(t-1), 0, \text{cap}_i(t))
\end{equation}
where \(\sigma \ge 0\) is the rate at which the fallback recovery capacity is restored. The desired recovery is:
\[
\text{desired\_recovery}_i(t) \]
\begin{equation}
= 
\begin{cases}
\mu \cdot A(t) \cdot \max(\text{EMA}_i(t) - UV_i(t), 0) & \text{EMA}_i(t) > 0 \\
0 & \text{otherwise}
\end{cases}
\end{equation}
where \( \mu \) is the recovery rate constant. The actual recovery is:
\begin{equation}
\text{recovery}_i(t) = \min\left( \text{desired\_recovery}_i(t), \text{max\_recovery}_i(t) \right)
\label{eq:lim_recov_mec}
\end{equation}
The utility value updates as:
\begin{equation}
UV_i(t+1) = UV_i(t) + \text{recovery}_i(t) + \kappa
\end{equation}
where \(\kappa = PUV_i(t) + \text{diffusion}_i(t)\) are other terms including P-UF inputs and diffusion effects. The EMS updates post-recovery.

\subsubsection{Modulation of P-UF Valence Outputs}

Anger may also modulate the valence outputs of P-UF components, reflecting the biological reality of reduced pain sensitivity during anger states (e.g., via adrenaline-driven mechanisms). The valence output is adjusted as:

\begin{equation}
v_j(t) = v_j^{\text{base}}(t) \cdot \left(1 - k_j \cdot \frac{\text{min}(A(t), A_\text{thres})}{A_{\text{thres}}}\right)
\label{eq:anger_modded_puf}
\end{equation}

where:
\begin{itemize}
    \item \( v_j^{\text{base}}(t) \) is the base valence derived from sensory input,
    \item \( k_j \) (\( 0 \leq k_j \leq 1 \)) is the modulation strength for component \( j \).
\end{itemize}

This modulation reduces the valence of P-UF components when anger is high, lessening their impact on S-UF perceptrons through the \( PUV_i(t) \) term. This further protects the current S-UF valence mapping and prioritizes threat response, mirroring adaptive survival behaviors.

\subsubsection{Local Anger Distribution}
Global anger \( A_p(t) \) maps to active perceptrons as a local anger scalar \( a_i(t) \). First, we introduce variable \(m_i(t)\) representing global anger effect on perceptron \(i\) at timestep \(t\):
\begin{equation}
m_i(t) = \lambda \cdot w_A(i) \cdot A(t) \cdot e_i(t)
\end{equation}
where \(\lambda\) (\(0 < \lambda \le 1\)) is an anger sensitivity coefficient, \( w_A(i) \ \) is the evolutionary-tuned weight for anger mapping, and \( e_i(t) \) is the eligibility trace.
In addition, negative valence is also mapped onto active perceptrons, resulting in the finalized update function:
\begin{equation}
UV_i(t+1) = \text{clip}(UV_i(t) - m_i(t) + \text{recovery}_i(t) + \kappa, -L_i, L_i)
\label{eq:anger_neg_uv}
\end{equation}
Note that the balance between \(-m_i(t)\) and \(\text{recovery}_i(t)\) relies on parameter tuning. \(UV_i(t) > 0\) biases the system towards approach behaviors, while \(UV_i(t) < 0\) favors avoidance.

\subsubsection{Diffusion of Local Anger}
Local anger diffuses to active, connected perceptrons:
\begin{equation}
\text{diffusion}_{a_i}(t) = f_s(i, t) \cdot \bigl(d_i(g_a, t) - l_i(g_a, t) \bigr)
\label{eq:diffusion_anger}
\end{equation}
where \(g_a(i, t) = a_i(t)\), and
\begin{equation}
a_i(t+1) = a_i(t) + m_i(t) + \text{diffusion}_{a_i}(t)
\label{eq:local_anger_scalar}
\end{equation}
where \( \eta \) (\( 0 \le \eta \le 1 \)) is the loss rate.

Local anger is clipped:
\begin{equation}
a_i(t+1) = \min\left(a_i(t+1), L_i\right)
\label{eq:p_anger_lim}
\end{equation}
with \( L_i \) as the perceptron-specific limit.

\subsubsection{Threat Neutralization}
When the threat triggering anger is resolved (e.g. through external changes or internal reassessment) or enough time has passed, anger \(A(t)\) naturally decreases as \(\text{Loss}(t)\) diminishes due to baseline recalibration via the EMA. This reduction restores \(\beta_\text{effective}(t)\) toward \(\beta\), potentially allowing extinction of positive S-UF components if the threat persists without resolution. The DE module may further modulate anger- such as by adjusting global anger \(A(t)\), adjusting local anger \(a_i(t)\), or redirecting it to alternative targets. Under assumption (ii), valence can travel via diffusion along active perceptron chains that reflect conscious reasoning, such as linguistic reframing, based on the DE module’s situational assessment. However, the specific mechanisms of the DE-driven modulation such as appraisal or attention are complex and context-dependent, warranting exploration in future work.

This formulation ensures anger designates targets for suppression, temporarily protects positive valence, and maintains system stability by capping recovery. Note that anger and frustration are both modeled by this system, though the cause of the initial decrease in positive valence acting as the trigger may be different. For anger, the decrease of a positive S-UF component is caused by a negative S-UF or P-UF activation, inducing valence cancellation. For frustration, the decrease is caused by repeated activation of the positive S-UF component without the anticipated reward from the P-UF.

\subsection{Toy Simulation}
\begin{figure}[!t]
  \centering
  \includegraphics[width=0.9\columnwidth]{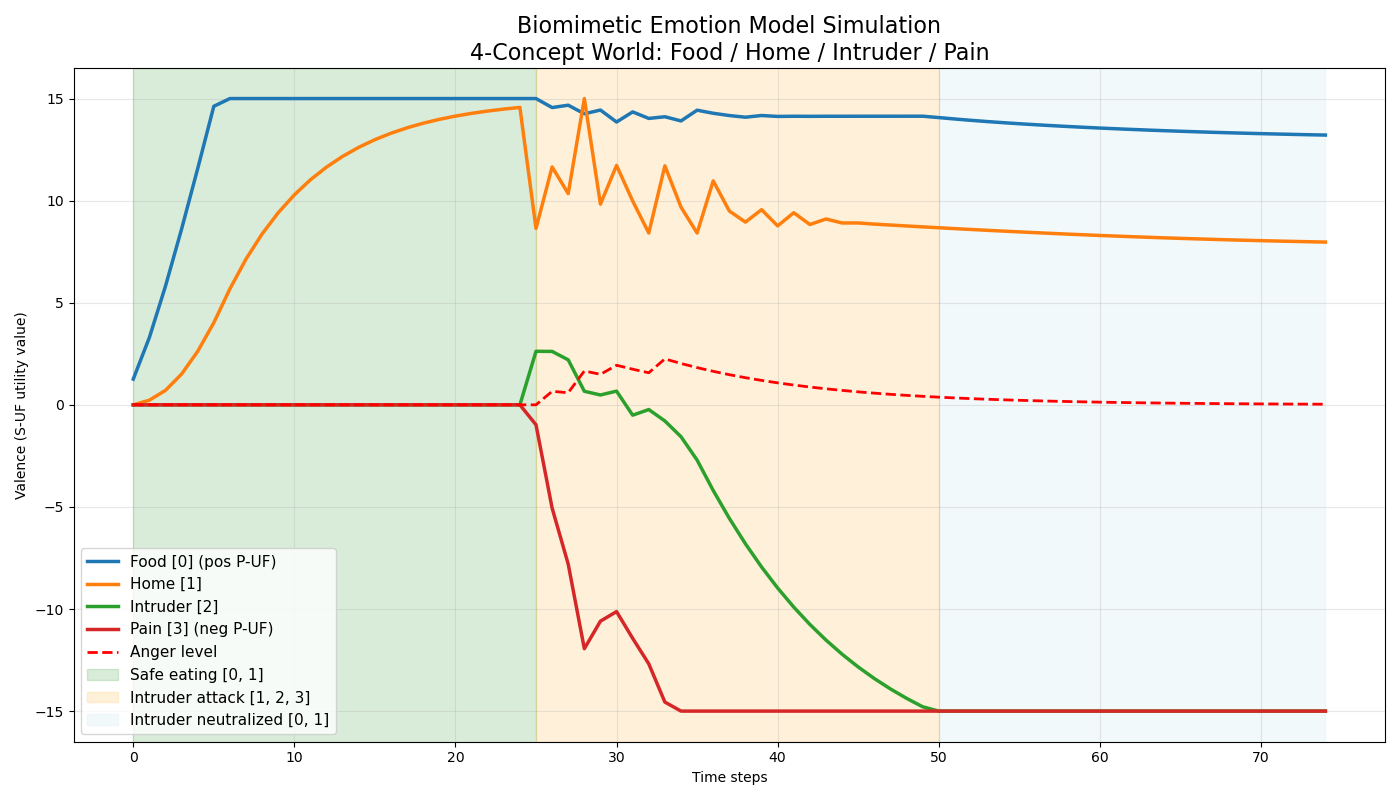}
  \caption{Output of the 4-concept simulation showing valence trajectories under safe eating, intruder attack, and recovery phases.}
  \label{fig:sim_valence}
\end{figure}

To illustrate the core valence dynamics (diffusion, depletion, anger protection), we implemented a minimal 4-concept toy world (food, home, intruder, pain) in Python. The code is available in a public repository \cite{SimpleEmotionSim}, and the output is shown in Fig. \ref{fig:sim_valence}. Food and pain are both tied to positive and negative P-UF components, respectively. The simulation has three phases. The first and third phases have [`home', `food'] concepts active, while the second has [`home', `intruder', `pain'] active. Notice that the first UV peak of `home' due to valence recovery is higher than the peak before the threat phase, which can be taken advantage of given correct timing (e.g. teasing). Also notice that while the `food' P-UF is active in the third phase, there is no valence recovery. This is because sufficient local anger is still active, stopping P-UF influence.

\section{Scenarios}

The framework encompasses a variety of scenarios, spanning from direct implementations of the outlined formulations to more complex emergent behaviors. To illustrate these possibilities, we examine selected examples, categorized into two groups. ``Direct behavior'' scenarios reflect immediate applications of the UF and anger dynamics not already addressed earlier in the text, where emotional responses align closely with the model's core formulations. In contrast, ``Emergent behavior'' scenarios arise from the interplay of multiple components, revealing how the system generates complex emotional phenomena not explicitly encoded in the formulations.

\subsection{Direct behavior}

We begin with direct behavior scenarios, aiming to collectively engage the core formulations of our model.

\subsubsection{P-UF driven needs (Hunger)}
Initially, autonomic processes detect the need for food, triggering positive valence gustatory P-UF components associated with taste and satiety. These components amplify the motivation towards S-UFs via persistent connections, guiding the individual toward seeking nourishment.

If the system remains unsatisfied for an extended period, a global negative valence is introduced through interoceptive P-UF components of starvation, possibly biased toward evolutionary defined areas likely unrelated to seeking of sustenance. This shift increases the motivational contrast between approach and avoidance behaviors, compelling the individual to avoid unrelated behavior and intensify their search for food. If the system continues to remain unsatisfied, the positive S-UF components will deplete (Eq. (\ref{eq:diffusion_valence})) from repeated activation without P-UF restoration, leaving only avoidance of negative-valenced hunger as a motivational drive.

During consumption, P-UFs linked with taste and satiety are activated, effectively mapping valence onto active perceptrons (Eq. (\ref{eq:puv})) to reinforce positive experiences. Once satiated, P-UF components utilize persistent connections to deactivate approach motivations without altering the established valence mappings, leaving them intact for future reference.

\subsubsection{Negative S-UF (Internal avoidance)}

S-UF components within this framework can be triggered in various ways—either internally, externally, or through a combination of both (partial activation described by Eq. (\ref{eq:partial_activation})). When activation occurs solely from within, such as through thoughts or memories, these components can be subject to internal avoidance mechanisms. This might manifest as behaviors like normalcy bias, where individuals downplay potential threats to maintain a sense of stability, or procrastination, where action is delayed to avoid discomfort. In contrast, when S-UF components are activated by external stimuli—such as a real-time event or sensory input—avoiding them becomes far more challenging, as the external trigger demands immediate attention and cannot simply be ignored or suppressed internally. Over time, with sufficient experience, the DE module may adapt by learning to regulate emotions more strategically. For instance, it could deliberately activate a negative S-UF component when necessary—perhaps to spur action or confront a problem—driven by the reinforcement of a better long-term outcome. This outcome might even be tied to a positive S-UF concept, such as the satisfaction of achieving a goal, which encapsulates the expected net positive result and motivates the modulation process.

In extreme cases, such as profound trauma, the dynamics of internal avoidance can take a more intense form. Here, strong negative valence may overwhelm the system, mapping onto all active nodes during the traumatic experience. This includes not only the perceptrons heavily targeted by the negative P-UF components but also less directly involved ones, influenced by the transfer of negative valence through S-UF connections from more weighted targets (\(w_j(i)\) term) and the sheer volume of emotional intensity. As a result, even normal bodily sensations—like breathing or heartbeat—can become entangled in this outsized negative S-UF component. When this occurs, the individual may experience dissociation, a state where the mind avoids physical awareness itself as a protective response to the overwhelming negativity, illustrating how the framework can account for severe emotional disruptions.

\subsubsection{Evolutionarily primed S-UF (Psychopathy and Sadism)}

The flexibility of evolutionarily primed S-UF components, while adaptive for social bonding, can turn maladaptive when valence mapping deviates. In psychopathy, social cues like happiness and distress signals fail to acquire positive or negative valence, possibly due to impaired S-UF learning from weak P-UF weighting (\(w_j(i)\)) or low perceptron valence limits (\(L_i\)), leaving empathy and guilt stunted—e.g. a happy or pained expression remains neutral rather than rewarding or aversive, respectively. Conversely, sadism sees positive P-UF outputs, such as pleasure from coercive extraction of value, reinforce S-UF concepts tied to the suffering of others, flipping distress into reward. These potential aberrations- observed in reduced amygdala response \cite{Birbaumer_2005} or aberrant reward circuits \cite{Buckholtz_2010}
-highlight how primed flexibility can skew emotional outcomes.

\subsubsection{Threatened positive S-UF (Maternal protectiveness)}

Maternal protectiveness serves as an exemplar of targeted anger within our framework. During childbirth, positive P-UF components map significant valence onto an S-UF component representing the offspring during the imprinting process. If the amount of positive valence committed is sufficiently high, system-wide positive valence may be exhausted and contribute to postpartum depression. A potential threat (e.g., a predator) activates a negative S-UF, prompting an appraisal process described by Lazarus \cite{Lazarus_1991} by the DE module. If the situation is deemed a threat to the offspring, perceptron chain activations link the negative valence source to the S-UF concept representing the offspring (assumption ii and Eq. (\ref{eq:diffusion_valence})), leading to negative valence transfer and erosion of positive valence. This results in anger when the loss of positive valence exceeds a threshold. If harm to the offspring has already occurred, the valence cancellation may occur even without appraisal calculation. Anger dynamics protect the offspring S-UF by engaging in limited valence recovery (Eq. (\ref{eq:lim_recov_mec})) and reducing diffusion (Eq. (\ref{eq:beta_effective})). Global anger maps onto active concepts as local anger scalars (Eq. (\ref{eq:local_anger_scalar})) and normal negative valence (\(m_i(t)\) term in Eq. (\ref{eq:anger_neg_uv})) to define and target the threat. P-UF influence diminishes (Eq. (\ref{eq:anger_modded_puf})), prioritizing threat response over personal utility. Successful neutralization restores valence as global anger dissipates (Eqs. (\ref{eq:global_anger}), (\ref{eq:lim_recov_mec})); persistent threats risk offspring S-UF depletion.

\subsubsection{Positive S-UF depletion (Grief)}

The stages of grief as outlined in the Kubler-Ross Model \cite{Kubler_2005} can be understood in the context of an extinction cycle of a very strong positive S-UF component. Initially, an external event triggers the actual loss of a positive mapped concept, but the associated internal S-UF component persists. This persistence underlies the first stage of denial, where individuals internally avoid acknowledging the loss, attempting to preserve the positive valence through internal avoidance.

When avoidance fails, the system transitions to anger. Here, the significant positive valence decrease triggers anger (Eq. (\ref{eq:global_anger})), along with its protective mechanisms—consistent with the \textit{Anger Dynamics} subsection—to defend the positive S-UF component from depletion. It ramps up arousal and allocates resources to sustain the component’s valence, despite the absence of P-UF reinforcement. This heightened state reflects an emotional "fight" against the loss.

The third stage of bargaining reflects a diminished intensity of both global anger (normalization of baseline in Eq. (\ref{eq:global_anger})) and local anger scalars (diffusion losses via continued stimulation, Eq. (\ref{eq:diffusion_anger})), where the DE module continues to seek methods to counteract or mitigate the impending valence cancellation from perceptrons previously targeted by anger's suppressive mapping (\(-m_i(t)\) term in Eq. (\ref{eq:anger_neg_uv})). As local anger fades, this effort shifts from external threat suppression to internal reconfiguration or reframing to preserve positive valence.

In the fourth stage, depression emerges as prolonged failure exhausts the system’s capacity to sustain anger and heightened arousal. With protective mechanisms depleted, the positive S-UF component continues to lose valence through transfer inefficiencies and cancellation from anger's suppressive mapping (\(-m_i(t)\)). This erosion from lingering negativity underscores the emotional toll of the loss, marking a shift from active resistance to a state of depletion and despair.

Finally, in the fifth stage of acceptance, the positive source is either extinguished or greatly diminished, finalizing the recalibration of the S-UF component and a shift toward coping with the new reality.

\subsubsection{Positive S-UF depletion (Teasing)}

In a scenario involving teasing behavior, a positive S-UF component is activated in the individual being teased, inducing approach behavior. However, the associated P-UF component does not activate, and without reinforcement, the positive valence begins to deplete, signaling a potential loss. From the utility function’s perspective, this reflects either an incorrect predictor or a need for additional effort to secure the P-UF reward. To enhance motivation for approach, the S-UF valence is temporarily boosted via the limited recovery mechanism defined by Equation (\ref{eq:lim_recov_mec}), triggered by a low-level anger response to the valence drop. If the P-UF remains inactive, the S-UF’s valence depletes naturally through transfer inefficiencies, indicating unattainability, with residual anger lingering. Positive teasing capitalizes on this temporary boost to sustain engagement before frustration or full extinction sets in.

\subsubsection{Perceptron valence limit (Differing total valence capacity in concepts)}

Suppose there are limitations on the valence that a single perceptron can hold, as described by the \(L_i\) term in the formulations section. Consequently, concepts requiring more perceptrons to define would have a greater overall capacity for valence. If the complexity of a concept is generally proportional to the number of perceptrons needed to represent it, as implied by assumption (ii), then complex, multi-faceted concepts could accumulate higher levels of valence compared to simpler ones. This aligns with the intuition that simpler objects or entities, such as a toy dog, generally evoke less profound emotional responses than more complex entities, such as a real dog.


\subsubsection{Diffusion range limit (Coexisting valences)}

Coexisting valences could occur when opposite valence S-UF sources activate and persist, without the normal valence cancellation or anger reaction. This situation can arise from 1) sufficient distance such that the valence transfer between the two concepts via local diffusion is minimal (requires \(j \sim i\) as per Eq. (\ref{eq:diffusion_valence}), with transfer losses increasing in proportion to distance, further explored in the later "advertising" section), or 2) emotional modulation by the prefrontal cortex or DE module which suppresses cancellation to prioritize long-term goals.

\subsection{Emergent behavior}

This section explores emergent behaviors to illustrate the model’s versatility across cognitive, developmental, and social domains.

\subsubsection{Increased associations}

Suppose positive and negative activations of utility function components influence learning by strengthening associations between active concepts within range of the diffusion process, where functionally the association implies that activation of one concept increases the likelihood of activating another. In cases such as addiction, an aberrant positive P-UF is repeatedly activated under varying circumstances, with valence magnitudes beyond normal limits. Each new context introduces additional concurrent concepts, which become mapped with a positive valence (Eqs. (\ref{eq:puv}), (\ref{eq:uv_update})) and over time contribute to the expansion of the association network corresponding to the S-UF component. When the positive valence of this enlarged S-UF network is depleted—such as during drug rehabilitation—the underlying associations may persist. Consequently, even a single reactivation of the hijacked positive P-UF alongside a subset of the S-UF component can trigger activations across the associated network, amplifying the total mapped valence beyond that of the initial addiction formation. This behavior mirrors spontaneous recovery \cite{Bouton_2004}.

In contrast to maladaptive outcomes such as addiction, heightened associations within the network can also facilitate adaptive shifts in valence, particularly in the context of memory reevaluation. For concepts defined by robust internal connections, S-UF diffusion tends to consolidate valence into a unipolar state. For instance, a difficult memory linked to an extensive network of associations—culminating in a rewarding outcome—may gradually acquire a predominantly positive valence due to diffusion mechanics (Eq. (\ref{eq:diffusion_valence})). This process aligns with the psychological phenomenon of rosy retrospection \cite{Mitchell_1997}, where retrospective reevaluation casts past challenges in a favorable light, emphasizing the positive resolution over the initial hardship. Over many such experiences, the association network of positive, adaptive concepts increases in size. Repeated positive resolutions expand the network of adaptive associations, much like the persistent networks in addiction. Unlike addiction, however, this persistence may strengthen resilience against future hardships and setbacks.

\subsubsection{Language as valence diffusion conduit}

In humans, relevant components of the P-UF likely places significant emphasis on the language module. This focus suggests that language acts as a key conduit through which S-UF valence can extend beyond immediate sensory inputs and permeate more abstract concepts. By leveraging linguistic pathways, valence originating from concrete experiences can influence higher-level ideas.

\subsubsection{Abstract Emotions}

Abstract forms of utility, such as moral standards or societal hierarchies, can emerge from early social interactions. For instance, consider a child who experiences positive valence—stemming from tangible rewards like food or play—in association with maternal attention and approval. This pairing leads to the formation of a S-UF component linked to social feedback, reflecting the child’s learned emotional response to such interactions. Through ongoing social conditioning, this S-UF component can extend its influence into the linguistic domain, diffusing onto specific verbal labels and creating a corresponding S-UF component tied to language.

In this framework, language for humans serves as a critical conduit for emotional valence, enabling its spread from abstract concepts to concrete experiences. For example, the term "courage" might acquire positive valence through repeated association with societal approval, such as praise for brave actions, and transfer valence to representations of "courage" in other modalities. This process anchors abstract labels in real-world significance, reinforcing their emotional weight due to the perceptron-level valence limit discussed earlier.

\subsubsection{Music}

Language serves as a high-traffic conduit for valence, channeling emotional utility through auditory perceptrons to S-UF components, such as those tied to social approval or disapproval. Music, with its syntactic structure of patterns and resolutions akin to linguistic syntax \cite{Patel_2003}, engages concepts adjacent to language in the auditory domain. Due to the diffusion mechanics (Eq. (\ref{eq:diffusion_valence})), valence flow from language-related S-UF components accumulates in these activated and adjacent non-linguistic perceptrons. For example, a resolving chord progression may evoke positive valence by mimicking the closure of a spoken sentence, diffusing utility to concepts of comfort or reward. This process mirrors music’s emotional impact, where auditory patterns trigger valence transfer independent of explicit meaning.

\subsubsection{Humor}

Humor arises from dynamic interactions between concepts represented by perceptrons within the S-UF. Within this framework, neutral concepts—those lacking inherent positive or negative valence—can coexist alongside high-valence concepts but remain disconnected due to an absence of prior simultaneous activation. When an external stimulus activates both a neutral concept and a high-valence one concurrently, the high-valence concept can exert influence, facilitating a transfer of valence to the neutral one. If this transfer involves positive valence, the positive S-UF expands, enhancing the overall utility in a way that is perceived as beneficial; however, if the transfer carries negative valence, the result is less favorable.

The outcome of valence transfer between positive and negative S-UF components hinges on their relative strengths. When negative valence overshadows positive valence, the positive source may be perceived as being threatened, by default triggering anger. In contrast, when positive valence prevails over negative valence, the result is joy; no corrective action is needed, though the stimulation of the negative S-UF component might still manifest physiologically during the extinction process, such as in tears of joy.

Humor often emerges specifically when a neutral concept is suddenly flooded with positive valence, producing a delightful and unexpected emotional shift. This process underscores humor’s reliance on unexpected connections that abruptly alter emotional value, creating a surprising shift that elicits amusement. Similarly, exploratory behaviors aimed at broadening a positive S-UF component align with the concept of play, reflecting a related yet distinct mechanism where positive valence fuels engagement without necessitating a specific threat or resolution. Collectively, these interactions demonstrate how the framework links emotional experiences like humor and play to the underlying mechanics of valence transfer, offering a structured explanation for their emergence and impact.

\subsubsection{Advertising (Less is more)}

In the realm of emotional communication, valence transfer efficiency is a critical concept for understanding how messages evoke desired emotional responses. It refers to the degree to which a message’s emotional value- its valence- is preserved as it travels through a sequence of words or phrases, termed a linguistic chain. This principle is especially relevant in advertising and propaganda, where the primary goal is often to trigger a strong, undiluted emotional reaction in the audience, then transfer it to a target concept, whether a commercial brand or political concept.

Within this model, an external stimulus, such as appealing auditory and visual stimuli, activates S-UF components with high valence. The valence can then be redirected to a target concept via the language conduit, such as through the use of an advertising slogan. Each word or phrase in a linguistic chain serves as a node, and with every additional node, there is a risk of transfer losses and diffusion onto active nodes irrelevant to the target concept.

Per the diffusion process described by Equation (\ref{eq:diffusion_valence}) with \( \eta \) (0 \(\leq \eta \leq\) 1) as the transfer loss factor, valence diminishes with each step. A short chain (e.g., three perceptrons for ``Just Do It'') retains high efficiency (e.g., \( (1-\eta)^3 \approx 0.7 \) if \( \eta = 0.1 \)), delivering a strong \( UV_i(t) \) (e.g., +7.3 if the initial valence is +10) to drive approach behavior, such as brand loyalty. A longer chain (e.g., nine perceptrons for ``You should just go ahead and do it now'') incurs greater losses (e.g., \( (1-\eta)^{9} \approx 0.4 \)), reducing the final valence (e.g., +3.9) and diluting its emotional clarity.

To address this challenge, advertisers and propagandists frequently rely on short, punchy slogans. These concise linguistic chains minimize transfer losses, delivering the emotional impact with immediacy and force. Shorter messages demand less cognitive effort to process, enabling them to sidestep complex reasoning and directly stimulate emotional centers in the brain, such as the amygdala. This direct pathway enhances the message’s ability to provoke the intended emotional response with maximum potency, highlighting why brevity is a cornerstone of emotionally resonant messaging.

\subsubsection{Propaganda (Exploitation of anger)}

Anger’s protective mechanisms, as modeled in this framework, can be exploited by bad actors to manipulate populations into harmful actions. An example is Nazi Germany in the 1940s, where mass media falsely accused the Polish of exterminating ethnic Germans. Though untrue, this lie sparked widespread anger among believers, leading to Poles being early camp targets \cite{Evans_2009}. Within the model, such manipulation exploits two processes from the \textit{Anger Dynamics} formulations: S-UF diffusion decrease and P-UF valence modulation.

The S-UF diffusion decrease defined by Equation (\ref{eq:beta_effective}) limits valence spread as anger \( A(t) \) surges. Propaganda’s short, potent claims (e.g., ``Poles are killing Germans'') spike \( A(t) \), shrinking \( \beta_{\text{effective}}(t) \) and confining positive valence to threatened S-UF components (e.g., personal safety, national pride). This restricts diffusion to counteracting concepts like skepticism or evidence, locking individuals into an intense, defensive mindset primed for manipulation.

Meanwhile, P-UF valence modulation given by Equation (\ref{eq:anger_modded_puf}) dampens innate valence as anger peaks. Inflated \( A(t) \) from false threats reduces sensitivity to reality-based P-UF inputs, such as those from peaceful interactions with Polish neighbors. This blinds individuals to disconfirming evidence, sustaining anger for bad actors to exploit.

This dynamic reveals a perilous trap: unchecked anger, especially when fueled by lies, can be weaponized. This vulnerability suggests a counter-strategy: individuals can counter this by slowing anger’s onset, verifying claims personally before committing. Such caution restores S-UF diffusion from counteracting concepts and preserves P-UF’s balanced input, aligning emotions with truth rather than deceit.

\section{Conclusion}

As stated from the start, our objective in this work is not to provide an exhaustive explanation of emotions, but rather to offer a structured framework for understanding emotional mechanisms. By integrating basic and constructionist theories into a unified model through functionalist principles, we have created a relatively formal system by abstracting aspects of emotional processes and underlying physical mechanisms while maintaining their essential dynamics. This abstraction allows us to reason about complex emotional phenomena in a vastly simplified system, including isolating specific mechanisms and examining their roles in shaping emotional states, postulating possible mechanisms given environmental forces, and reconciling devised interactions within the model with observed behavior in actual biological organisms.

Moving forward, this framework could serve as a foundation for further research into how sensory inputs shape learned emotions or how social contexts modulate valence transfer. Expanding this model may yield deeper insights into the mechanisms governing emotional experiences and inform strategies for managing emotional challenges in individuals and groups.

\section{Acknowledgments}
\noindent The author is grateful to Drs. Torunsky and Knauz for valuable criticism and suggestions that improved the materials presented in this paper. Much thanks also to Dr. Pelz, who provided important feedback on the mathematical formulations.

\bibliographystyle{IEEEtran}
\bibliography{main}

\end{document}